\documentclass[twocolumn,aps,pra,amsmath,amssymb]{revtex4-1}
\usepackage[latin9]{inputenc}
\usepackage{amstext}
\usepackage{amssymb}
\usepackage{graphicx}
\usepackage{esint}

\makeatletter

\newcommand*\LyXThinSpace{\,\hspace{0pt}}

\usepackage{epsfig}
\usepackage{bm}
\usepackage{dcolumn}
\usepackage{color}

\makeatother

\begin{document}

\title{Many-body localization in XY spin chains with long-range interactions:
An exact diagonalization study}

\author{Sebastian Schiffer}

\author{Jia Wang}

\author{Xia-Ji Liu}

\author{Hui Hu}

\affiliation{Centre for Quantum and Optical Science, Swinburne University of Technology,
Melbourne, Victoria 3122, Australia}

\date{\today}
\begin{abstract}
We investigate the transition from the many-body localized phase to
the ergodic thermalized phase at an infinite temperature in an $XY$
spin chain with $L$ spins, which experiences power-law decaying interactions
in the form of $V_{ij}\propto1/\left|i-j\right|^{\alpha}$ ($i,j=1,\cdots,L$)
and a random transverse field. By performing large-scale exact diagonalization
for the chain size up to $L=18$, we systematically analyze the energy
gap statistics, half-chain entanglement entropy, and uncertainty of
the entanglement entropy of the system at different interaction exponents
$\alpha$. The finite-size critical scaling allows us to determine
the critical disorder strength $W_{c}$ and critical exponent $\nu$
at the many-body localization phase transition, as a function of the
interaction exponent $\alpha$ in the limit $L\rightarrow\infty$.
We find that both $W_{c}$ and $\nu$ diverge when $\alpha$ decreases
to a critical power $\alpha_{c}\simeq1.16\pm0.17$, indicating the
absence of many-body localization for $\alpha<\alpha_{c}$. Our result
is useful to resolve the contradiction on the critical power found
in two previous studies, $\alpha_{c}=3/2$ from scaling argument in
Phys. Rev. B \textbf{92}, 104428 (2015) and $\alpha_{c}\approx1$
from quantum dynamics simulation in Phys. Rev. A \textbf{99}, 033610
(2019).
\end{abstract}
\maketitle

\section{Introduction}

After Anderson published his famous paper on single-particle localization
in 1958 \cite{Anderson1958}, he and his collaborator Fleishman considered
the possibility that the insulation properties of the single-particle
localization also hold in the presence of inter-particle interactions
\cite{Fleishman1980}. It took a long time to eventually show that
this possibility is true for interacting many-body systems \cite{Gornyi2005,Basko2006}.
Many-body localization (MBL) since then became a flourishing research
frontier that attracts intense attention from different fields of
physics. For recent reviews, see, for example, Refs. \cite{Nandkishore2015,Altman2015,Abanin2019}. 

MBL systems defy the laws of standard quantum statistics, by explicitly
violating the eigenstate thermalization hypothesis \cite{Nandkishore2015,Altman2015,Abanin2019}
and preventing themselves from thermal equilibration. This makes them
interesting to study, for the purpose of obtaining a new understanding
of quantum physics. These systems are also fascinating because of
their unique features to block all transport phenomena. The fact that
MBL preserves the initial states of the system makes it important
for practical applications such as storage systems for qubits in a
quantum computer \cite{Nandkishore2015,Altman2015,Abanin2019,Wang2018}.

To date, there are a number of techniques developed to understand
MBL, including analytic calculations \cite{Feigelman2010}, numerically
exact diagonalization \cite{Wang2018,Oganesyan2007,Pal2010,Luitz2015,Hu2016,Khemani2017},
renormalization group approaches such as the excited-state real-space
renormalization group and density matrix renormalization groups (DMRG
and its time-dependent version tDMRG) \cite{Monthus2010,Pekker2014,Potter2015,Lim2016,Dumitrescu2017,Zhang2018},
and perturbation methods such as Born approximation \cite{Lev2014}
and self-consistent theories \cite{Prelovsek2017}. Recently, a renormalization
flow technique, namely the Wegner-Wilson flow renormalization, has
also been applied to investigate the MBL phase transition \cite{Pekker2017}.

Even though MBL has now been investigated for quite a while and general
consensus is slowly gained \cite{Nandkishore2015,Altman2015,Abanin2019},
the understanding of such an intriguing phenomenon in some many-body
systems remains as a challenge. In particular, there is a debate concerning
the possibility of MBL phase transition in disordered spin chains
with long-range power-law decaying interactions (i.e., $V_{ij}\propto1/\left|i-j\right|^{\alpha}$
for two spins at site $i$ and at site $j$) \cite{Burin2006,Yao2014,Burin2015,Hauke2015,Li2016,Nandkishore2017,Tikhonov2018,SafaviNaini2019,Roy2019}.
For a sufficiently large interaction exponent $\alpha\rightarrow\infty$,
the interaction is essentially short-range, for which the existence
of MBL is widely accepted \cite{Luitz2015,Imbrie2016}. However, in
general, one may anticipate that MBL will cease to exist when the
interaction exponent is smaller than a threshold, $\alpha<\alpha_{c}$,
where $\alpha_{c}$ depends on the dimensionality $d$ and also on
the type of the system \cite{Burin2006}. 

Actually, in the case of Anderson localization with single-particle
power-law hopping terms, an old argument by Anderson establishes $\alpha_{c}=d$,
based on the breakdown of perturbative expansion \cite{Fleishman1980}.
This argument was recently generalized to interacting spin systems,
by considering resonant spin-pair excitations that lead to the threshold
$\alpha_{c}=3d/2$ for an $XY$ chain \cite{Burin2015} and $\alpha_{c}=2d$
\cite{Tikhonov2018} for a Heisenberg chain. These nice predictions,
unfortunately, have not been rigorously examined by extensive numerical
calculations. This seems necessary, as the breakdown of perturbation
expansion is not equivalent to the breakdown of localization \cite{Nandkishore2017}. 

In a recent quantum dynamics study in one dimension (1D), growth of
entanglement entropy and quantum Fisher information were simulated
\cite{SafaviNaini2019}. While the results for the Heisenberg spin
chain are consistent with the predicted critical interaction exponent
$\alpha_{c}=2$, the results for the $XY$ spin chain indicate $\alpha_{c}\approx1$,
smaller than the predicted threshold $\alpha_{c}=3/2$. The disagreement
for the $XY$ chain suggests that MBL in such a system needs more
stringent numerical tests. This is a timely task considering its experimental
relevance. Most recently, a disordered $XY$ chain with power-law
interactions has been successfully engineered by using strings of
up to $20$ trapped $^{40}$Ca$^{+}$ ions \cite{Brydges2019}.

To resolve the discrepancy between the perturbative argument \cite{Burin2015}
and the dynamics simulation \cite{SafaviNaini2019}, here we present
an extensive finite-size scaling study of a disordered 1D $XY$ spin
chain with power-law decaying interactions for system sizes up to
$L=18$ spins, by using large-scale exact diagonalization (ED) and
using the standard MBL indicators such as the energy gap statistics,
half-chain entanglement entropy, and uncertainty of the entanglement
entropy. These indicators were previously used to \emph{convincingly}
establish the MBL transition in spin chains with nearest-neighbor
(short-range) interactions and with the number of spins up to $22$
\cite{Luitz2015}. The largest number of spins simulated in this work
($L=18$) is somehow smaller, since our Hamiltonian matrix with long-range
$XY$ interactions becomes much denser than those in the case of short-range
nearest-neighbor interactions. Nevertheless, our size is larger the
typical size of $L=14$ taken in the earlier ED study for disordered
$XY$ chains \cite{Burin2015}, allowing us to unambiguously examining
MBL at different interaction exponents $\alpha$ and hence to reliably
determine the critical interaction exponent $\alpha_{c}$. It is interesting
to note that, our size is very close to the size of the experimentally
engineered $XY$ spin chain (i.e., $L=20$) \cite{Brydges2019}. Therefore,
the results obtained in this work might be useful for future experimental
investigations.

\begin{figure}[t]
\centering{}\includegraphics[width=0.5\textwidth]{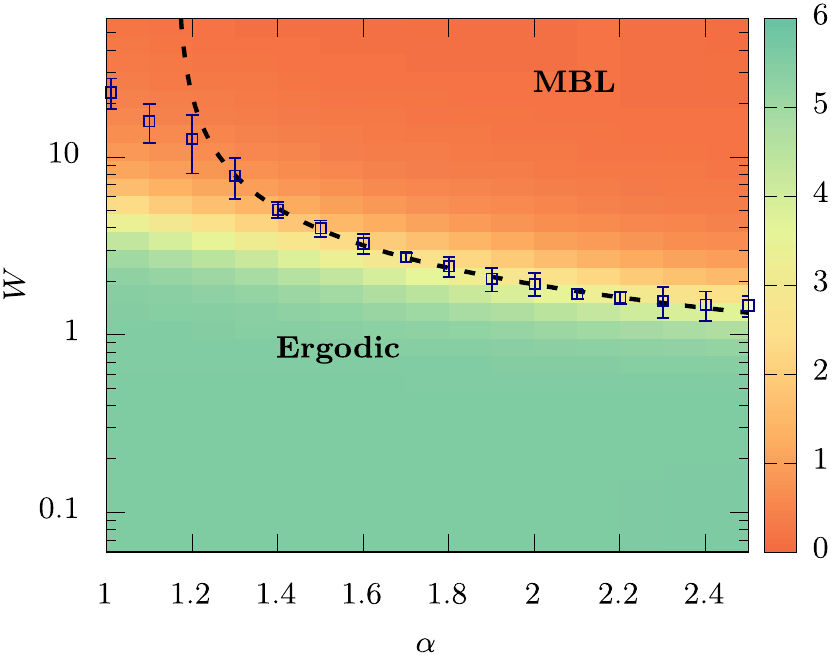}
\caption{\label{fig1_PhaseDiagramL18} Contour plot of the half-chain entanglement
entropy $S_{E}$ as functions of the interaction exponent $\alpha$
(in a linear scale) and the disorder strength $W$ (in a logarithmic
scale), for an $XY$ spin chain with $L=18$ spins. The red region
marks the MBL phase, whereas in the green region the system is in
the ergodic phase. The blue squares indicate the critical disorder
strength $W_{c}(\alpha)$ in the thermodynamic limit obtained from
finite-size scaling analysis, which seems to diverge with decreasing
$\alpha$. The blue square at $\alpha=1.0$ is slightly displaced,
in order to clearly show the error bar. A power-law fit to $W_{c}(\alpha)$,
as described in Sec. IV and Appendix B, leads to the determination
of the phase boundary (dashed black line) and a critical interaction
exponent $\alpha_{c}\simeq1.16\pm0.17$. }
\end{figure}

Our main results are briefly summarized in Fig. \ref{fig1_PhaseDiagramL18},
which reports the half-chain entanglement entropy ($S_{E}$) of the
$XY$ spin chain with $L=18$ spins, as functions of the interaction
exponent $\alpha$ and disorder strength $W$. We can easily identify
an ergodic thermalized phase at weak disorder and an MBL phase when
the disorder is sufficiently strong. The finite-size scaling at different
interaction exponents enables us to determine the critical disorder
strength $W_{c}(\alpha)$ in the thermodynamic limit $L\rightarrow\infty$.
We find that $W_{c}(\alpha)$ increases rapidly as we decrease the
interaction exponent $\alpha$ down to $1.0$. At the same time, the
uncertainty of $W_{c}(\alpha)$ indicated by the finite-size scaling
also dramatically increases. By using a power-law fit to $W_{c}(\alpha)$,
we extract a critical interaction exponent $\alpha_{c}\simeq1.16\pm0.17$,
which is consistent with the dynamics simulation \cite{SafaviNaini2019}.
We do not find any singular behavior of the critical disorder strength
$W_{c}$ at $\alpha=3/2$, which is in tension with the prediction
by the perturbative argument based on the consideration of resonant
spin-pair excitations \cite{Burin2015}.

\section{Disordered 1D XY spin chains}

We consider an $XY$ chain with total $L$ spins in a random transverse
field with power-law decaying interactions, described by the model
Hamiltonian \cite{SafaviNaini2019,Brydges2019}, 
\begin{equation}
\mathcal{\hat{H}}=\sum_{1\leq i<j\leq L}\frac{J_{0}}{\left|j-i\right|^{\alpha}}\left(\hat{\sigma}_{i}^{+}\hat{\sigma}_{j}^{-}+\hat{\sigma}_{i}^{-}\hat{\sigma}_{j}^{+}\right)+\sum_{i=1}^{L}h_{i}\hat{\sigma}_{i}^{z}\label{eq:xy_ham}
\end{equation}
where $\sigma_{i}^{\pm}=(\hat{\sigma}_{i}^{x}\pm i\hat{\sigma}_{i}^{y}$)/2
and $\hat{\sigma}_{i}^{z}$ are the Pauli matrices at site $i$ and
we take an exchange interaction strength $J_{0}=1$ as the units of
energy. The interaction exponent $\alpha>0$ characterizes the range
of interactions. In the case of an infinitely large $\alpha\rightarrow\infty$,
we recover the short-range nearest-neighbor interaction. The random
transverse field $h_{i}$ is uniformly distributed in the interval
$[-W,+W]$. This model has conserved $z$-component of total spin
operator $\hat{S}_{z}=\sum_{i}\hat{\sigma}_{i}^{z}$, i.e., $[\mathcal{\hat{H}},\hat{S}_{z}]=0.$
As a result, we consider the sector $\hat{S}_{z}=0$.

To examine MBL at infinite temperature, we consider the many-body
energy levels near \emph{zero} energy. This is obvious for short-range
interactions, where the energy levels distribute symmetrically with
respect to zero energy. In our case with long-range interactions,
we may define an average energy $\epsilon$ of the system at temperature
$T$, $\epsilon=\text{Tr}(\mathcal{\hat{H}}e^{-\beta\mathcal{\hat{H}}})/\text{Tr}(e^{-\beta\mathcal{\hat{H}}}),$
where $\beta=1/(k_{B}T)$. By evaluating $\epsilon$ using ED at different
interaction exponents, we find that $\epsilon$ always are very close
to zero when we increase the temperature to infinity.

In our simulations, the model Hamiltonian is solved by ED and the
$50$ eigenstates with energy closest to zero energy $\epsilon=0$
are chosen. For each point evaluated at certain disorder strength
$W$ and interaction exponent $\alpha$, various MBL indicators are
calculated for this set of eigenstates and are averaged over $10^{3}$
different disorder realizations ($600$ for the largest system size
$L=18$). 

\begin{figure}[t]
\centering{}\includegraphics[width=0.5\textwidth]{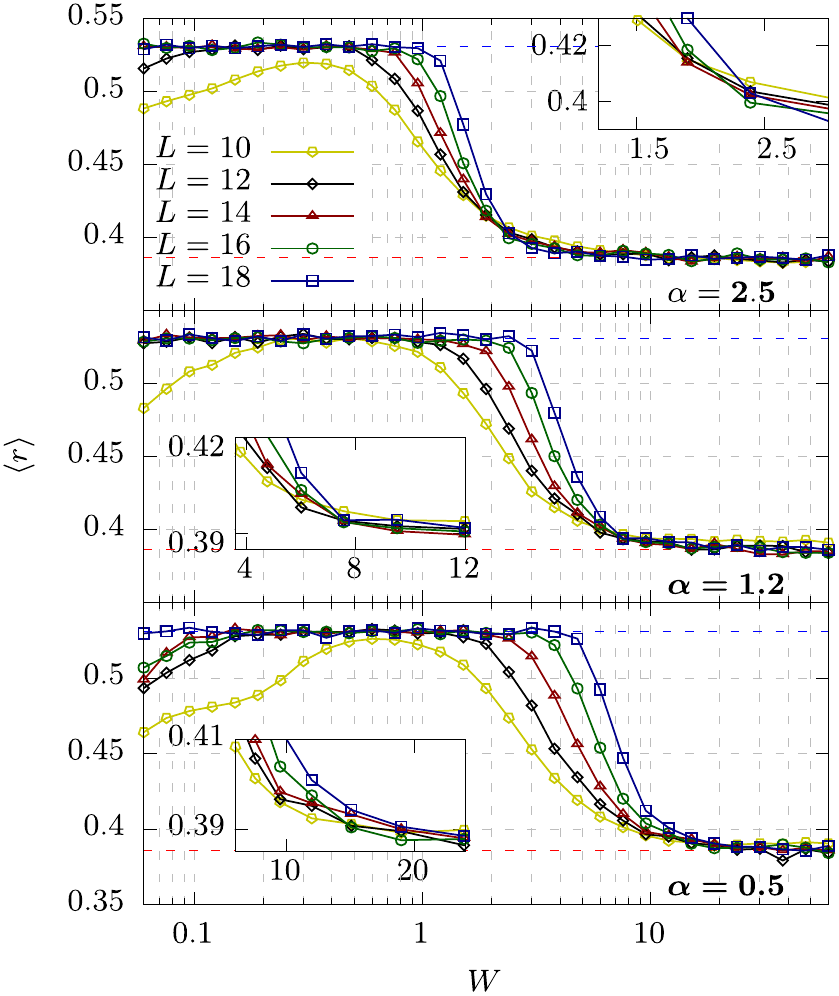}
\caption{\label{fig2_EnergyGapStatistics} Averaged ratio of successive gaps
$\langle r\rangle$ for $\alpha=2.5$ (top), $\alpha=1.2$ (middle)
and $\alpha=0.5$ (bottom), at different spin chain lengths. The dashed
blue line indicates the thermal limit, $\left\langle r\right\rangle _{\textrm{GOE}}\simeq0.5307$,
whereas the red dashed line shows the localized limit, $\left\langle r\right\rangle _{\textrm{Poisson}}\simeq0.3863$.
The insets highlight the area near the phase transition, where the
crossing point between curves with different chain lengths is anticipated
to appear.}
\end{figure}

\section{MBL indicators with finite spins}

The first convenient MBL indicator is the energy gap statistics characterized
by the averaged ratio of successive gaps \cite{Oganesyan2007,Pal2010},
$\langle r\rangle=\left\langle \min\{\delta_{n+1},\delta_{n}\}/\max\{\delta_{n+1},\delta_{n}\}\right\rangle $,
where $\delta_{n}\equiv E_{n+1}-E_{n}$ and $E_{n}$ is the energy
of the $n$th eigenstate closest to zero energy. This ratio takes
values between $\langle r\rangle_{\textrm{GOE}}\approx0.5307$ for
a Gaussian orthogonal ensemble (GOE) in the thermalized phase and
$\langle r\rangle_{\textrm{Possion}}=2\ln(2)-1\approx0.3863$ for
a Poisson distribution in the MBL phase. The disorder strength dependences
of $\left\langle r\right\rangle $ at $\alpha=2.5$, $1.2$ and $0.5$
are presented in Fig. \ref{fig2_EnergyGapStatistics} (from top to
bottom), together with the two limiting values indicated by the blue
and red dashed lines, respectively.

In all the subplots, the gap statistics are close to the GOE prediction
at weak disorder and approach the Poisson limit at sufficiently strong
disorder, indicating the possibility of a phase transition in between.
At an infinite system size, the phase transition would manifest itself
as a sudden jump in $\left\langle r\right\rangle $ from GOE to Poisson
limits at a certain critical disorder strength $W_{c}$. For the finite-size
system simulated in this work, we instead see a smooth crossover,
as anticipated. The possible existence of a phase transition may be
characterized by the crossing point between curves corresponding to
$L$ and $L-2$ spins, which approaches $W_{c}$ as $L$ increases
and hence provides a lower-bound estimation of $W_{c}$. As shown
in the insets, for $\alpha=2.5$ we find a clear drift of the cross
point at about $W\sim2.5$. For $\alpha=1.2$, the shift of the cross
point becomes difficult to identify. The situation for $\alpha=0.5$
is even worse. The crossing point seems to lie at much stronger disorder
strength around $W\sim30$, where the quality of the data does not
allow us to determine possible crossing points. Overall, we find that
the MBL transition, if it exists, becomes increasingly difficult to
occur as the interaction exponent $\alpha$ decreases.

\begin{figure}[t]
\centering{}\includegraphics[width=0.5\textwidth]{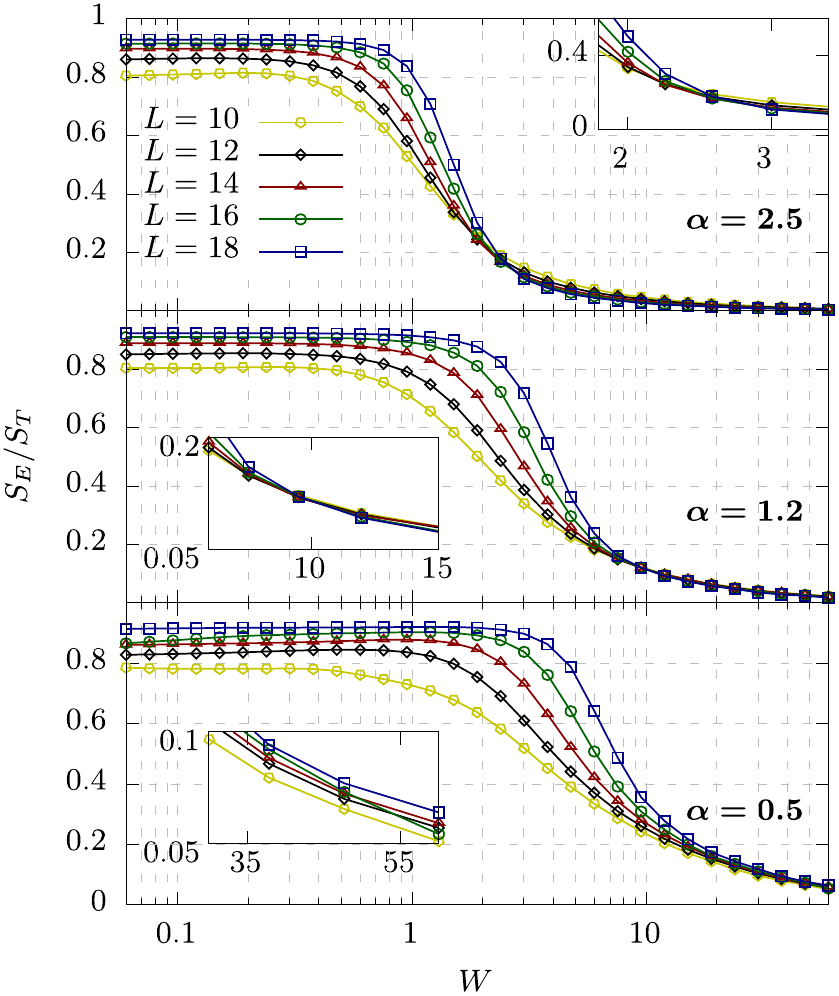} \caption{\label{fig3_EntanglementEntropy} Half-chain entanglement entropy
$S_{E}$ for $\alpha=2.5$ (top), $\alpha=1.2$ (middle) and $\alpha=0.5$
(bottom), normalized by $S_{T}\equiv(L\ln2-1)/2$, as a function of
the disorder strength at different chain sizes. The insets highlight
the region that MBL may occur.}
\end{figure}

The crossing points may also be seen in the \emph{normalized} half-chain
entanglement entropy $S_{E}$ at finite spins, which is another MBL
diagnostics \cite{Pal2010}. At finite $L$, the normalization is
provided by the Page value $S_{T}\equiv(L\ln2-1)/2$ for a random
pure state \cite{Page1993}. In the thermal phase $S_{E}$ is expected
to reach $S_{T}$, exhibiting a volume law; while deep in the MBL
phase it would follow an area law and become independent of the system
size, making $S_{E}/S_{T}$ vanishingly small for large system size.
These two limiting behaviors are clearly shown in Fig. \ref{fig3_EntanglementEntropy},
where we present $S_{E}/S_{T}$ as a function of the disorder strength
at different system sizes for $\alpha=2.5$ (top), $\alpha=1.2$ (middle)
and $\alpha=0.5$ (bottom). In the former two cases, we can clearly
identify a crossing point at $W\sim2.5$ and $W\sim12$, respectively,
similar to what is found in the energy gap statistics. In sharp contrast,
for $\alpha=0.5$ with increasing disorder strength, the curves of
the normalized entropy at different $L$ roughly decrease in parallel.
This makes it impossible to locate a meaningful crossing point.

\begin{figure}[t]
\centering{}\includegraphics[width=0.5\textwidth]{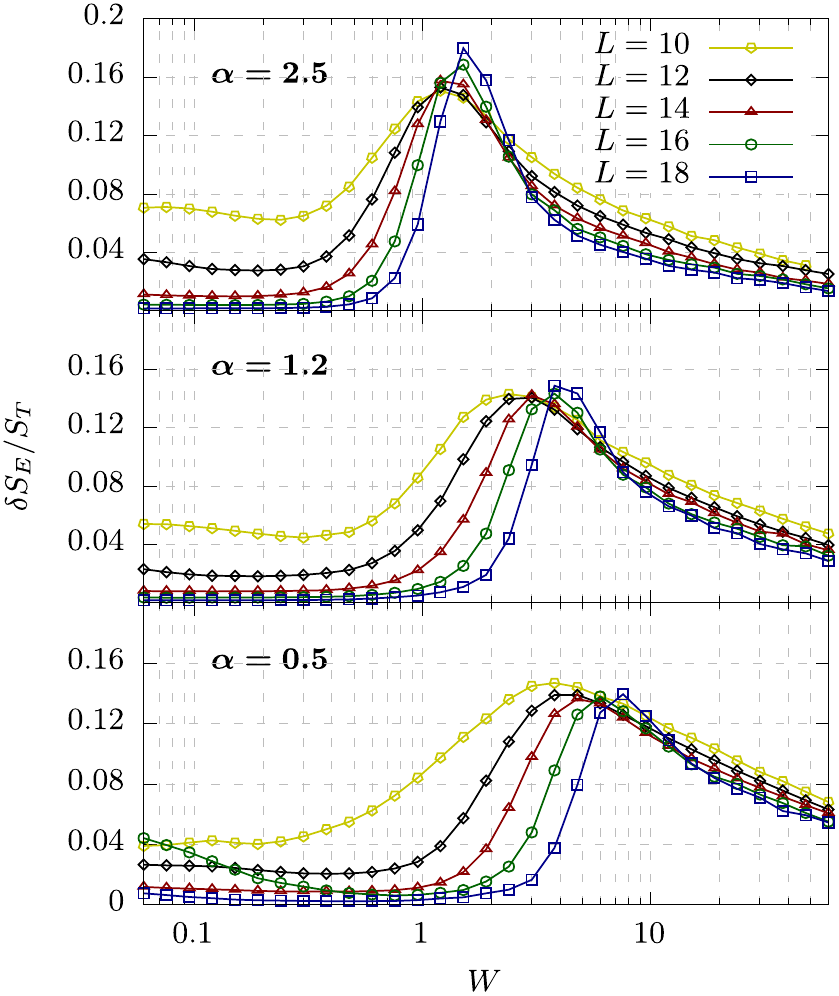}
\caption{\label{fig4_DeltaEE} Uncertainty of the entanglement entropy $\delta S_{E}$
for $\alpha=2.5$ (top), $\alpha=1.2$ (middle) and $\alpha=0.5$
(bottom), divided by the Page value $S_{T}\equiv(L\ln2-1)/2$, as
a function of the disorder strength at different chain sizes $L$.
For all the interaction exponents, with increasing $L$ the peak positions
of the curves move towards stronger disorder.}
\end{figure}

As the final MBL indicator, we check the uncertainty of the entanglement
entropy $\delta S_{E}$, which peaks at the thermal to MBL transition
but tends to vanish both in the deep thermal and MBL phases \cite{Kjall2014}.
Figure \ref{fig4_DeltaEE} reports the size-dependence of $\delta S_{E}$
(in units of the Page value $S_{T}$) at the three interaction exponents
$\alpha=2.5$ (top), $\alpha=1.2$ (middle) and $\alpha=0.5$ (bottom).
Two observations are worth noting. First, for the largest interaction
exponent $\alpha=2.5$, the peak value of $\delta S_{E}/S_{T}$ grows
with $L$, indicating that $\delta S_{E}$ increases super-linearly
with $L$ for the system sizes under consideration. Similar growth
was previously observed for short-range interaction models \cite{Khemani2017},
where MBL transition is known to occur. As we decrease $\alpha$,
the growth in $\delta S_{E}/S_{T}$ becomes much weaker at $\alpha=1.2$
and stops completely at $\alpha=0.5$. On the other hand, for all
the three interaction exponents, the peak position of $\delta S_{E}/S_{T}$
moves to the right towards the side of strong disorder. Naïvely, we
may interpret the peak position as the size-dependent critical strength
$W_{c}(L)$, as it essentially plays the same role of the crossing
point that we find in $\left\langle r\right\rangle $ and $S_{E}/S_{T}$.
At $\alpha=2.5$, the shift of the peak position slows down with increasing
$L$, suggesting the saturation to a finite critical disorder strength
in the thermodynamic limit $L\rightarrow\infty$. On the contrary,
we find that the movement of the peak position at $\alpha=0.5$ is
much faster. As $L$ increases the peak position scales at least linearly
in $L$, implying an infinitely large critical disorder strength in
the thermodynamic limit and hence the absence of the MBL transition.
For more details, we refer to Appendix A.

From the three MBL indicators, we may conclude the existence and absence
of the MBL phase transition at large ($\alpha=2.5$) and small interaction
exponents ($\alpha=0.5$), respectively. The case of an intermediate
interaction exponent, i.e., $\alpha=1.2$, turns out to be marginal
and requires further exploration.

\begin{figure}[t]
\centering{}\includegraphics[width=0.5\textwidth]{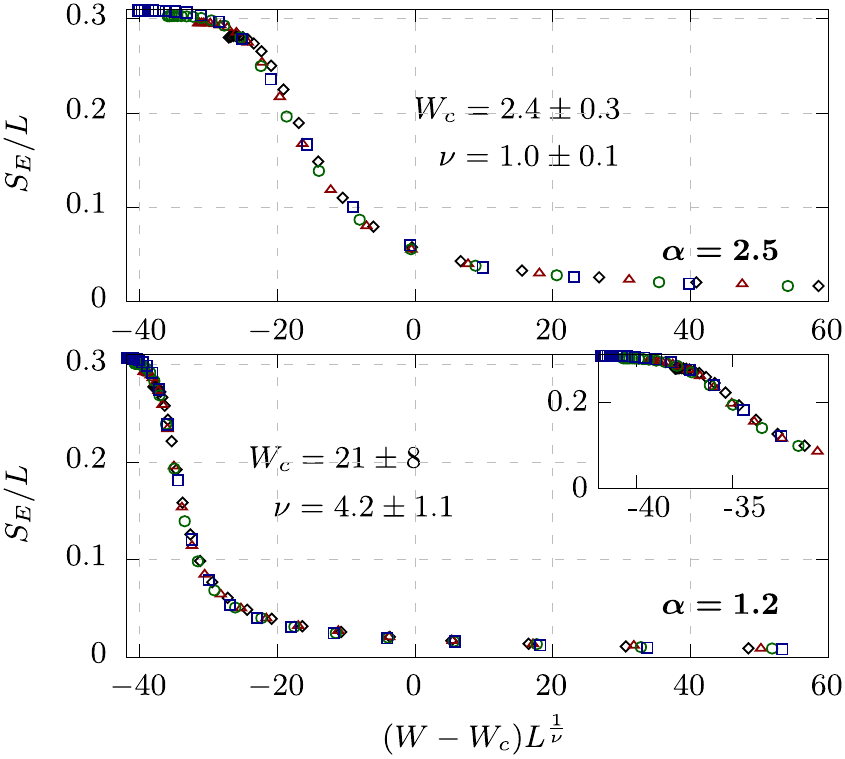}
\caption{\label{fig5_FiniteSizeScalingSE} Finite-size critical scaling collapse
for the data of the half-chain entanglement entropy $S_{E}/L$ at
$\alpha=2.5$ (top) and $\alpha=1.2$ (bottom). The data at different
chain lengths are shown by blue squares ($L=18$), green circles ($L=16$),
red triangles ($L=14$) and black diamonds ($L=12$). The inset in
the bottom panel blows up the thermalized phase. As the interaction
exponent $\alpha$ decreases from $2.5$ to $1.2$, the universal
curve obtained from the critical scaling changes significantly. At
$\alpha=2.5$, we find that $W_{c}=2.4\pm0.3$ and $\nu=1.0\pm0.1$;
while at $\alpha=1.2$, $W_{c}=21\pm8$ and $\nu=4.2\pm1.1$. The
critical exponent $\nu$ turns out to be very different in the two
cases.}
\end{figure}

\section{Finite-size scaling}

We thus consider the finite-size scaling properties of the data for
the three MBL indicators. Focusing on the normalized half-chain entanglement
entropy $S_{E}/L$, near the MBL transition (i.e., $W\sim W_{c}$),
the data might be fit to the scaling form \cite{Luitz2015},

\begin{equation}
S_{E}(L,W)=L\tilde{f}\left[\frac{L}{\xi\left(W\right)}\right]=Lf\left[\left(W-W_{c}\right)L^{1/\nu}\right],
\end{equation}
where $\xi(W)\propto\left|W-W_{c}\right|^{-\nu}$ is the correlation
length near the transition and $\nu$ is the critical exponent. As
shown in Fig. \ref{fig5_FiniteSizeScalingSE}, we find that for both
$\alpha=2.5$ and $\alpha=1.2$, the normalized entropy data sets
at different chain lengths collapse nicely onto each other. A similar
scaling collapse is also observed for the spectral gap statistics
(see Appendix B). The excellent scaling collapse might be viewed as
a convincing confirmation of the existence of an MBL phase transition,
particularly for the case of $\alpha=1.2$, where the naïve trace
of the crossing points in both $\left\langle r\right\rangle $ and
$S_{E}/S_{T}$ fails to draw conclusions. We note, however, that the
scaling collapse at $\alpha=1.2$ comes with \emph{large} errors for
the critical disorder strength $W_{c}$ and critical exponent $\nu$,
both of which are used as fitting parameters in the data collapse.
This is an important feature we shall discuss in the following.

\begin{figure}[t]
\centering{}\includegraphics[width=0.5\textwidth]{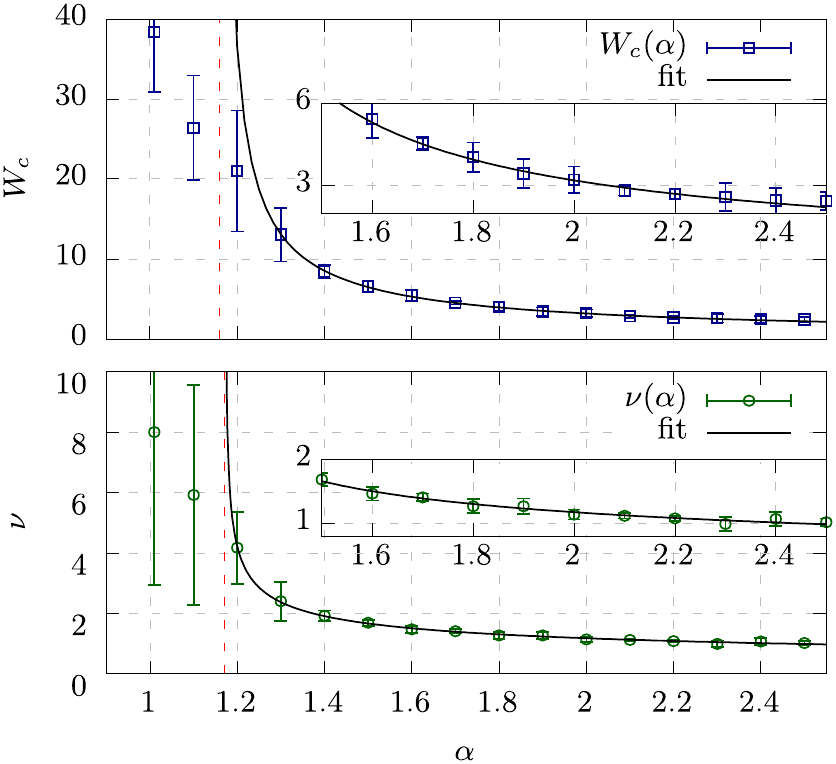}
\caption{\label{fig6_WcAndNu} Critical disorder strength $W_{c}$ (top) and
critical exponent $\nu$ (bottom), obtained from the finite-size scaling,
as a function of the interaction exponent $\alpha$. They are fitted
by a power-law formalism (see text) to yield a critical interaction
exponent $\alpha_{c}\approx1.16$, which is indicated by the vertical
dashed red lines. The insets highlight $W_{c}$ and $\nu$ at large
interaction exponent $\alpha$.}
\end{figure}

The same finite-size scaling analysis is applied to multiple data
sets of $S_{E}/L$ for $\alpha$ in the range $[1.0,2.5]$. The resulting
critical disorder strength and critical exponent are reported in Fig.
\ref{fig6_WcAndNu}. With decreasing $\alpha$ close to 1.0, it is
easy to see that both parameters start to diverge, along with a dramatically
increasing uncertainty. This is a strong indication of the existence
of a critical interaction exponent $\alpha_{c}$, below which the
system is unable to be many-body localized even for an arbitrarily
strong disorder. 

To determine $\alpha_{c}$, we fit $W_{c}(\alpha)$ and $\nu(\alpha)$
by using the following power-law formalism,

\begin{equation}
\eta\left(\alpha\right)=A_{\eta}\left(\alpha-\alpha_{c,\eta}\right){}^{-\gamma_{\eta}},\label{eq:FittingFormalism}
\end{equation}
where $\eta$ stands for either $W_{c}$ or $\nu$. As the system
size in our simulation is still relatively small, the divergence in
$W_{c}(\alpha)$ and $\nu(\alpha)$ can not fully manifest itself
when $\alpha$ is close to $\alpha_{c}$. To overcome this subtlety,
we impose a low-bound $\alpha_{f}$ and only the data at $\alpha>\alpha_{f}$
are selected for the curve fitting. We choose $\alpha_{f}$ in such
a way that the fitting errors for the fitting parameters $A_{\eta}$,
$\gamma_{\eta}$ and $\alpha_{c,\eta}$ are minimized (see Appendix
B). 

For the critical disorder strength, the best fit then leads to $\alpha_{f}=1.3$,
$\gamma_{W_{c}}=0.78\pm0.06$, and $\alpha_{c,W_{c}}=1.16\pm0.03$.
For the critical exponent, we instead obtain $\alpha_{f}=1.2$, $\gamma_{\nu}=0.38\pm0.02$,
and $\alpha_{c,\nu}=1.17\pm0.01$. The different exponents $\gamma$
found in the two fittings should not be taken seriously, since in
principle there is no constraint for their equality. It is remarkable
that the two fittings lead to essentially the same critical interaction
exponent $\alpha_{c}\simeq1.16$. We note that, here the error for
$\alpha_{c}$ only counts for the numerical error of the curve fitting.
It does not \emph{fully} include the large uncertainty in $W_{c}(\alpha)$
and $\nu(\alpha)$ near $\alpha_{c}$ that we emphasized earlier.
To take them into account in a more reasonable way, we use the bootstrap
resampling (see Appendix C). We find $\alpha_{c,W_{c}}=1.16\pm0.17$
and $\alpha_{c,\nu}=1.17\pm0.14$. Conservatively, we therefore conclude,
\begin{equation}
\alpha_{c}=1.16\pm0.17.
\end{equation}
This is the central result of our work.

A few remarks are now in order. First, the critical interaction exponent
obtained in the above could be useful to resolve the discrepancy in
$\alpha_{c}$, predicted by the perturbative argument based on the
resonant spin-pair excitations \cite{Burin2015} or calculated from
the dynamics simulation for the growth of entanglement entropy and
for the imbalance \cite{SafaviNaini2019}. The latter (with $\alpha_{c}\approx1$)
is supported by our ED study. The good agreement on $\alpha_{c}$
suggests that the two numerical calculations complement each other.
As the quantum dynamics simulation can access longer spin chains (i.e.,
$L=30$ for the $XY$ model and $L=40$ for the Heisenberg model in
\cite{SafaviNaini2019}) than ED, we believe that our finite-size
scaling analysis could be reliable and robust, against future ED studies
with larger $L$, considering the rapidly increasing capacity in computation.
There is some tension between our result $\alpha_{c}\simeq1.16$ and
the prediction $\alpha_{c}=3/2$ from the perturbative argument \cite{Burin2015}.
Our finite-size scaling analysis does not show any singular behavior
at $\alpha=1.5$. As we tune the interaction exponent $\alpha$ across
$1.5$, both the critical disorder strength $W_{c}$ and the critical
exponent $\nu$ change rather smoothly, with small uncertainties comparable
to those at large $\alpha$ (i.e., at $\alpha=2.5$).

Second, unless at $\alpha\lesssim1.4$ the critical exponent $\nu(\alpha)$
determined from our finite-size scaling violates the rigorous Harris/CCFS/CLO
scaling bound that requires $\nu\geq2/d=2$ \cite{Harris1974,Chayes1986,Chandran2105}.
This is a well-known problem for the finite-size scaling analysis
of the MBL transition. For the MBL transition in models with short-range
interactions, the critical exponent $\nu$ extracted from finite-size
scaling is about $\nu=0.91\pm0.07$ or $0.80\pm0.04$ (for gap statistics
or entanglement entropy, up to $L=22$ \cite{Luitz2015}) or $\nu=1.09$
(for entanglement entropy, up to $L=18$ \cite{Khemani2017}), which
is much smaller than the prediction of $\nu=3.1\pm0.3$ obtained from
the real-space renormalization group analysis \cite{Dumitrescu2017,Zhang2018}.
The latter satisfies the Harris/CCFS/CLO bound. The small critical
exponent is argued due to the fact that the quenched randomness is
not fully manifested itself at the system sizes probed by ED studies,
as indicated by the super-linear increase in the uncertainty of the
entanglement entropy $\delta S_{E}(L)$ \cite{Khemani2017}. Our critical
exponent $\nu(\alpha)$ seems to recover the finite-size scaling result
$\nu\sim1$ for short-range models when $\alpha$ is large (i.e.,
$\alpha=2.5$), and we do observe the same super-linear increase of
$\delta S_{E}(L)$ with increasing $L$ (see Fig. \ref{fig4_DeltaEE},
the top panel). When we decrease $\alpha$ down to $\alpha_{c}$,
the critical exponent $\nu(\alpha)$ \emph{gradually} increases above
the Harris/CCFS/CLO bound, and at the same time, $\delta S_{E}(L)$
stops increasing super-linearly in a consistent way. The smooth change
makes us believe that the universality class of the MBL transition
with long-range interactions may belong to the same universal class
of short-range models \footnote{We note that, at the infinitely large interaction exponent ($\alpha\rightarrow\infty$),
our model describes non-interacting spinless fermions with on-site
disorder, which experience Anderson localization for an arbitrarily
small disorder strength.}. This anticipation might be confirmed by a real-space renormalization
group study for the long-range $XY$ model, if possible.

Finally, it is interesting to ask, what happens if we tune the interaction
exponent $\alpha$ across the critical power $\alpha_{c}$ at a given
strong disorder $W\gg1$? Here, the thermal to MBL transition is controlled
(or driven) by $\alpha$ and, if the transition is continuous we may
anticipate the correlation length $\xi$ diverges like $\xi(\alpha)=\left|\alpha-\alpha_{c}\right|^{-\nu}$
near the transition. As a result, the scaling law for the MBL indicators
takes the form, for example, $S_{E}(L,\alpha)/L=h[L^{1/\nu}\left(\alpha-\alpha_{c}\right)]$.
The finite-size scaling analysis may then give us an alternative way
to accurately determine the critical interaction exponent $\alpha_{c}$.

\section{Conclusions}

In conclusions, exact diagonalization of the model Hamiltonian for
a disordered $XY$ spin chain with long-range interactions has been
performed to address the many-body localization phase transition,
with the number of spins up to $18$. The energy gap statistics, half-chain
entanglement entropy and uncertainty of the entanglement entropy have
been calculated, as a function of the chain length $L$ for different
interaction exponents $\alpha$ that characterizes the range of interactions.
All the three many-body localization diagnostics, after finite-size
scaling analysis, suggest the existence of a critical interaction
exponent $\alpha_{c}\simeq1.16\pm0.17$, below which the many-body
localization disappears for arbitrary disorder strength in the thermodynamic
limit. This result may help to resolve the discrepancy on $\alpha_{c}$
found in two recent theoretical studies \cite{Burin2015,SafaviNaini2019}.
On the other hand, our result could also be useful for future experiments
on many-body localization, to be carried out with up to 20 trapped
$^{40}$Ca$^{+}$ ions \cite{Brydges2019} that simulate the disorder
$XY$ model with long-range interactions at the interaction exponent
$\alpha\sim1$. At this point, the small-size limitation of our exact
diagonalization study becomes less relevant.
\begin{acknowledgments}
This research was supported by the Australian Research Council's (ARC)
Discovery Programs, Grant No. DE180100592 (J.W.), Grant No. DP190100815
(J.W.), Grant No. FT140100003 (X.-J.L), Grant No. DP180102018 (X.-J.L),
and Grant No. DP170104008 (H.H.). All our numerical calculations were
performed using the new high-performance computing resources (OzSTAR)
at Swinburne University of Technology, Melbourne.
\end{acknowledgments}

\appendix

\section{Peak position in the uncertainty of the entanglement entropy}

\begin{figure}[b]
\centering{}\includegraphics[width=0.5\textwidth]{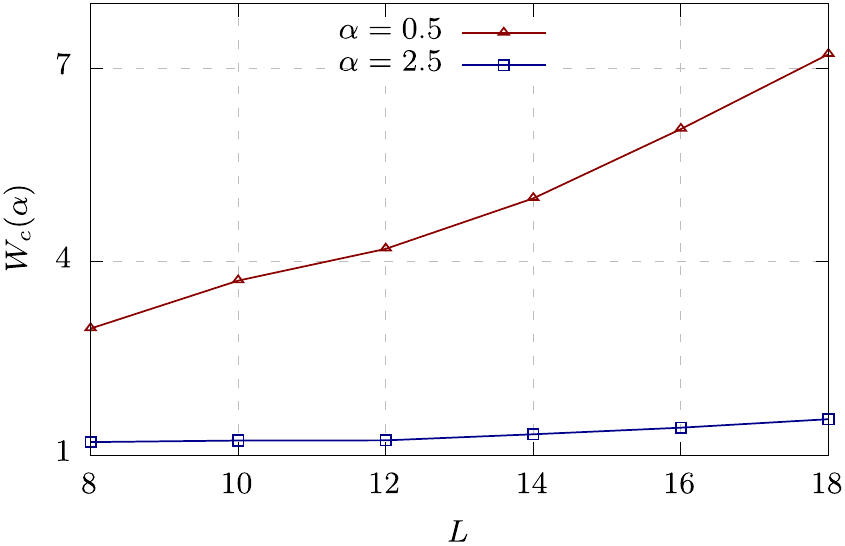} \caption{\label{fig7_WcL} The size dependence of the peak position of the
uncertainty of the entanglement entropy $\delta S_{E}$, at $\alpha=0.5$
and $\alpha=2.5$ as indicated.}
\end{figure}

Here we discuss in more detail the peaks appearing in the uncertainty
of the entanglement entropy $\delta S_{E}$. To reliably locate the
peak position, we sample more data points close to the maximum of
$\delta S_{E}$. At the same time, the number of different disorder
realizations is increased by a factor of $100$ for the chain lengths
up to $L=12$ and by a factor of ten for $L>12$.

Figure \ref{fig7_WcL} shows the peak position of $\delta S_{E}$
as a function of the system size $L$. This is treated as the critical
disorder strength $W_{c}(L)$ at the size $L$. It is easy to see
that $W_{c}(L)$ increases slowly and rapidly at $\alpha=2.5$ and
$\alpha=0.5$, respectively. The former is a sign of convergence towards
a finite critical disorder strength in the thermodynamic limit. For
the latter, we notice that $W_{c}(L)$ increases at least linearly
in $L$, which seems to rule out the possibility of many-body localization
when the system size becomes sufficiently large.

\begin{figure}
\centering{}\includegraphics[width=0.5\textwidth]{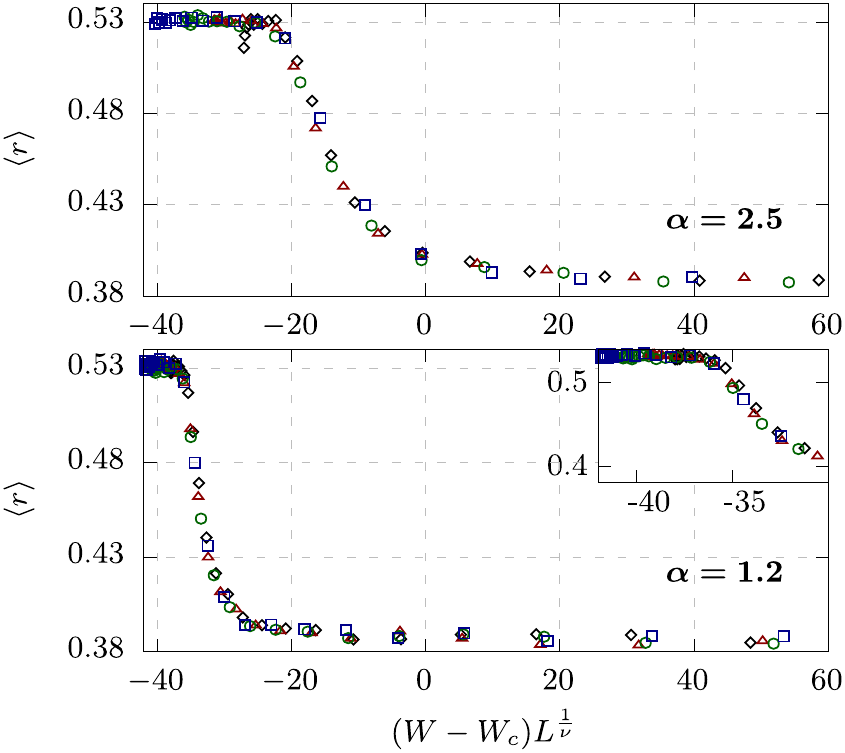}
\caption{\label{fig8_FiniteSizeScalingEnergyGap} The critical scaling collapse
for the data sets of the spectral gap statistics $\left\langle r\right\rangle $
at $\alpha=2.5$ (top) and $\alpha=1.2$ (bottom). The critical disorder
strength $W_{c}$ and the critical exponent $\nu$ are the same as
in Fig. \ref{fig5_FiniteSizeScalingSE}. }
\end{figure}

\section{More results on the finite-size scaling}

Figure \ref{fig8_FiniteSizeScalingEnergyGap} reports the critical
scaling collapse for the spectral gap statistics $\left\langle r\right\rangle $
at $\alpha=2.5$ (top) and $\alpha=1.2$ (bottom). To show the consistency
with the critical scaling for the entanglement entropy in Fig. \ref{fig5_FiniteSizeScalingSE},
we use the same critical disorder strength $W_{c}$ and the critical
exponent $\nu$. The collapse is very satisfactory, both for large
($\alpha=2.5$) and intermediate ($\alpha=1.2$) interaction exponents.

\begin{figure}[t]
\centering{}\includegraphics[width=0.5\textwidth]{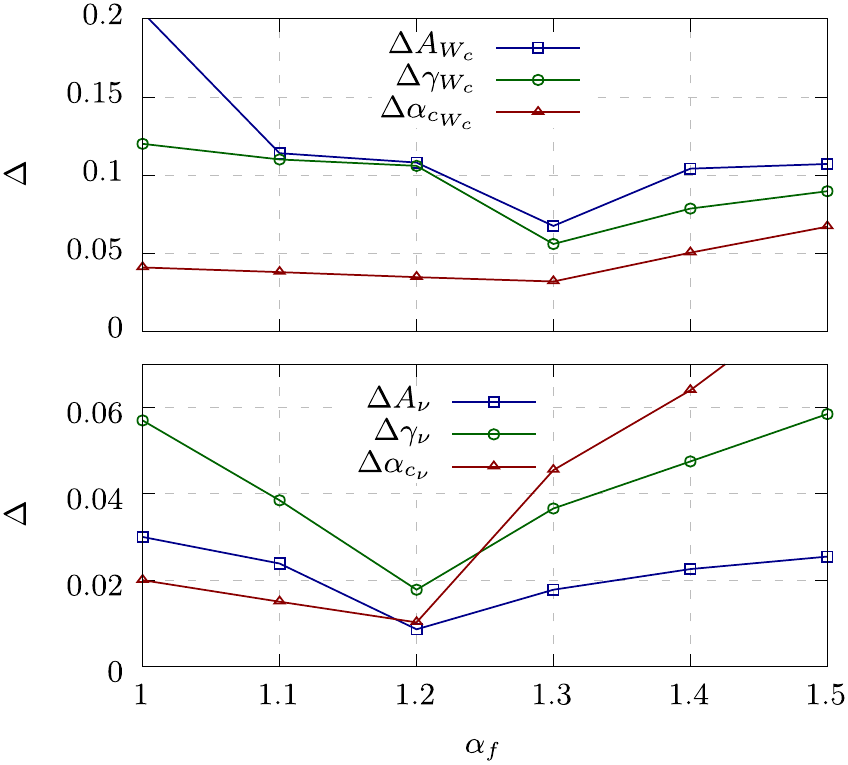}
\caption{\label{fig9_FittingErrors} The fitting errors of the fitting parameters
in Eq. (\ref{eq:FittingFormalism}) for the critical disorder strength
$W_{c}$ (upper panel) and for the critical exponent $\nu$ (lower
panel), as a function of the pre-selected $\alpha_{f}$.}
\end{figure}

Figure \ref{fig9_FittingErrors} shows the fitting error of the parameters
$A_{\eta}$, $\gamma_{\eta}$ and $\alpha_{c,\eta}$, obtained by
the curve fitting at a given pre-selected interaction exponent $\alpha_{f}$
for the data sets $W_{c}(\alpha)$ (upper panel) and $\nu(\alpha)$
(lower panel). There is a minimum for the fitting errors, occurring
at $\alpha_{f}=1.3$ for $W_{c}(\alpha)$ and at $\alpha_{f}=1.2$
for $\nu(\alpha)$. 

\section{Bootstrap resampling}

For a given set of data points $X=\{x_{i},\bar{y}_{i}\}$, we usually
fit them with a function $f(x;\vec{a})$ with fitting parameters $\vec{a}$
using standard softwares such as gnuplot and MATLAB, which do not
fully take into account the error $\delta y_{i}$ for the calculated
or measured value $\bar{y}_{i}$. This is partly due to the non-linearity
of the fitting function. To treat $\delta y_{i}$ in a more confident
way, a good strategy is \emph{resampling} the data points by assuming
noise $\delta y_{i}$. This is the so-called bootstrap resampling.

To implement the bootstrap resampling, for each data point $\{x,\bar{y}_{i}\}$,
we assume a normal distributed $y_{i}$ around the mean value $\bar{y}_{i}$
with standard deviation $\delta y_{i}$. We then generate a number
of new data sets $X_{k}=\{x_{i},y_{i}\}_{k}$, $k\in N_{\textrm{boot}}$,
where $N_{\textrm{boot}}$ is a sufficiently large integer. For each
generated data set $X_{k}$, we perform the fitting procedure and
obtain fitting parameters $\vec{a}_{k}$ with estimated errors. These
estimated errors lead to a standard deviation vector $\vec{s}_{\textrm{boot}}$,
which can be considered as a confident uncertainty of $\vec{a}$,
i.e., 
\begin{equation}
\Delta\vec{a}=\sqrt{\frac{N_{\textrm{boot}}}{N_{\textrm{boot}}-1}}\vec{s}_{\textrm{boot}}.
\end{equation}
In our calculations, we choose $N_{\textrm{boot}}\sim10^{5}$. The
bootstrap resampling leads to 
\begin{eqnarray}
\alpha_{c,W} & = & 1.16\pm0.17,\\
\gamma_{W} & = & 0.78\pm0.25,
\end{eqnarray}
for the data set of critical disorder strength with $\alpha_{f}=1.3$
and 
\begin{eqnarray}
\alpha_{c,\nu} & = & 1.17\pm0.14,\\
\gamma_{\nu} & = & 0.38\pm0.10,
\end{eqnarray}
for the data set of critical exponent with $\alpha_{f}=1.2$.

\end{document}